\documentclass[sigconf,anonymous=false,natbib=true]{acmart}
\usepackage[utf8]{inputenc}

\usepackage{booktabs} 
\usepackage{algorithm, algpseudocode}
\usepackage{amsmath}
\usepackage{graphics}
\usepackage{epsfig}
\usepackage{graphicx}
\usepackage{xcolor}
\usepackage{balance}
\usepackage{multirow}
\usepackage{mathrsfs}
\usepackage{acronym}
\usepackage{placeins}
\usepackage{color}
\usepackage[inline]{enumitem}
\usepackage{fancyhdr}
\usepackage{amsfonts}
\usepackage{svg}
\usepackage{threeparttable}
\usepackage{subfig}
\usepackage{float}
\usepackage{bbm}
\usepackage{array}
\usepackage{amssymb}
\usepackage{amsthm}
\usepackage[title]{appendix}

\AtBeginDocument{%
  \providecommand\BibTeX{{%
    \normalfont B\kern-0.5em{\scshape i\kern-0.25em b}\kern-0.8em\TeX}}}

\newcommand{\header}[1]{\vspace*{1mm}\noindent{\textbf{#1}}}

\copyrightyear{2024}
\acmYear{2024}
\setcopyright{rightsretained}
\acmConference[WSDM '24] {seventeenth ACM International Conference on Web Search and Data Mining}{March 4--8, 2024}{Mexico.}
\acmDOI{10.1145/3616855.3635848}
\acmBooktitle{seventeenth ACM International Conference on Web Search and Data Mining (WSDM '24), March 4--8, 2024, Mexico.}
\acmPrice{}
\begin{document}





\title[Debiasing Sequential Recommenders through DRO over System Exposure]{Debiasing Sequential Recommenders through Distributionally Robust Optimization over System Exposure}
\settopmatter{authorsperrow=4}
\author{Jiyuan Yang}
\email{jiyuan.yang@mail.sdu.edu.cn}
\affiliation{%
  \institution{Shandong University}
  \city{Qingdao}
  \country{China}
}

\author{Yue Ding}
\email{dingyue@sjtu.edu.cn}
\affiliation{%
  \institution{Shanghai Jiao Tong University}
  \city{Shanghai}
  \country{China}
}
\author{Yidan Wang}
\email{yidanwang@mail.sdu.edu.cn}
\affiliation{%
  \institution{Shandong University}
  \city{Qingdao}
  \country{China}
}
\author{Pengjie Ren}
\email{renpengjie@sdu.edu.cn}
\affiliation{%
  \institution{Shandong University}
  \city{Qingdao}
  \country{China}
}
\author{Zhumin Chen}
\email{chenzhumin@sdu.edu.cn}
\affiliation{%
  \institution{Shandong University}
  \city{Qingdao}
  \country{China}
}
\author{Fei Cai}
\email{caifei08@nudt.edu.cn}
\affiliation{%
  \institution{National University of Defense Technology}
  \city{Changsha}
  \country{China}
}
\author{Jun Ma}
\email{majun@sdu.edu.cn}
\affiliation{%
  \institution{Shandong University}
  \city{Qingdao}
  \country{China}
}

\author{Rui Zhang}
\email{rayteam@yeah.net}
\affiliation{%
  \institution{ruizhang.info}
  \city{Shenzhen}
  \country{China}
}
\author{Zhaochun Ren}
\email{z.ren@liacs.leidenuniv.nl}
\affiliation{%
  \institution{Leiden University}
  \city{Leiden}
  \country{Netherlands}
}
\author{Xin Xin}
\authornote{Corresponding author.}
\email{xinxin@sdu.edu.cn}
\affiliation{%
  \institution{Shandong University}
  \city{Qingdao}
  \country{China}
}


\begin{abstract}
Sequential recommendation (SR) models are typically trained on user-item interactions which are affected by the system exposure bias, leading to the user preference learned from the biased SR model not being fully consistent with the true user preference. Exposure bias refers to the fact that user interactions are dependent upon the partial items exposed to the user. Existing debiasing methods do not make full use of the system exposure data and suffer from sub-optimal recommendation performance and high variance.

In this paper, we propose to debias sequential recommenders through Distributionally Robust Optimization (DRO) over system exposure data.
The key idea is to utilize DRO to optimize the worst-case error over an uncertainty set to safeguard the model against distributional discrepancy caused by the exposure bias. 
The main challenge to apply DRO for exposure debiasing in sequential recommendation lies in how to construct the uncertainty set and avoid the overestimation of user preference on biased samples. Moreover, since the test set could also be affected by the exposure bias, how to evaluate the debiasing effect is also an open question.
To this end, we first introduce an exposure simulator trained upon the system exposure data to calculate the exposure distribution, which is then regarded as the nominal distribution to construct the uncertainty set of DRO. Then, we introduce a penalty to items with high exposure probability to avoid the overestimation of user preference for biased samples. 
Finally, we design a debiased self-normalized inverse propensity score (SNIPS) evaluator for evaluating the debiasing effect on the biased offline test set. 
We conduct extensive experiments on two real-world datasets to verify the effectiveness of the proposed methods. Experimental results demonstrate the superior exposure debiasing performance of proposed methods. 
Codes and data are available at \url{https://github.com/nancheng58/DebiasedSR_DRO}.
\end{abstract}
\maketitle
\acresetall

\section{Introduction}
\begin{figure}
    \captionsetup[subfloat]
    {}
    \subfloat[]{
        \label{intorduction-a}
        \includegraphics[width=0.4\linewidth]{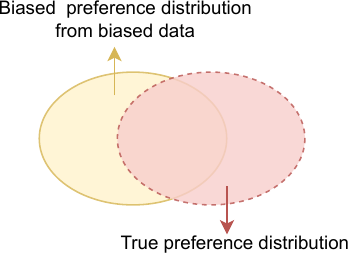}
    }
    \hspace{0.03cm}
    \vline
    \hspace{0.03cm}
    \subfloat[]{
        \label{intorduction-b}
        \includegraphics[width=0.52\linewidth]{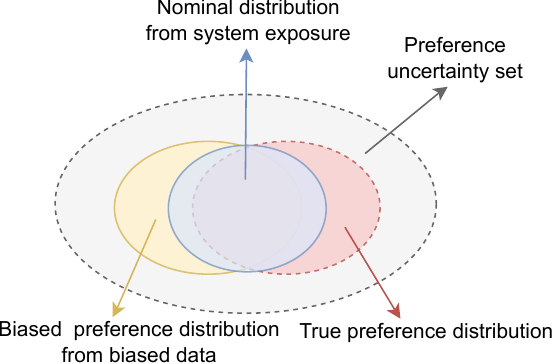}
    }
    \vspace{-0.1cm}
    \caption{Illustration of the user preference distribution gap (a) and our solution through DRO framework (b).}
    \vspace{-0.2cm}
    \label{fig:intorduction}
\end{figure}

Over the past decade, sequential recommendation (SR) has achieved great success in the field of recommender systems (RSs) to predict user future interests from sequential user-item interactions ~\cite{reinforce-e-commerce,nextitnet,tenrec,yaliWSDM23}. 
However, recent research has revealed that RSs face different intractable bias issues (e.g., popularity bias~\cite{DBLP:conf/sigir/ZhangF0WSL021}, position bias~\cite{DBLP:journals/corr/abs-1802-06565positionbias}, selection bias~\cite{wang2021combating}  and exposure bias~\cite{DBLP:conf/wsdm/SaitoYNSN20,10.1145/3306618.3314288,Yang2018unbiasevaluatuion}).
In particular, exposure bias occurs when user interactions are dependent upon the partial items exposed to the user, also known as the "previous model bias"~\cite{DBLP:conf/sigir/LiuCDHP020}.
Since the RS model is trained on user-item interaction data, the previously deployed RS model will dramatically impact the current one, and the exposed data in the current model is largely dependent on the previous recommendation policy, resulting in a biased feedback loop.
The RS may inadvertently exacerbate the bias since the agent is trained using biased interaction data, which leads to false predictions about user true preference. 
As shown in Fig. \ref{intorduction-a}, the user preference learned from exposure biased data is not fully consistent with the distribution of true user preference. Existing methods mainly utilize the inverse propensity score (IPS) to perform exposure debias \cite{DBLP:conf/wsdm/SaitoYNSN20, liu2023bounding, 10.1145/3306618.3314288, Khalil2022cloze, wang2022lantentconfounders, DBLP:conf/cikm/Xu0CDW22, dai2022generalized,wang2019doubly}. However, we argue that such kinds of methods do not make full use of system exposure data \cite{UserBehaviorLeakageTOIS23} and suffer from unbounded high variance \cite{DBLP:conf/www/ZhengGLHLJ21,DBLP:journals/jmlr/BottouPCCCPRSS13}. 

\noindent\textbf{Distributionally Robust Optimization.} In this paper, we propose to leverage \emph{Distributionally Robust Optimization} (DRO) \cite{DROsurvey,DBLP:journals/corr/abs-1911-08731,slowik2022distributionally, OODsurvey,yang2023generic,zhou2023distributionally} to perform debiased sequential recommendation. 
The motivation to utilize DRO is that DRO aims to optimize the worst-case error over an \emph{uncertainty set} to safeguard the model against potential distributional shifts within the uncertainty set \cite{OODsurvey}, which naturally fits the needs of debiasing sequential recommendation. Besides, DRO also provides more stable and robust model training compared with high variance IPS-based methods.
More specifically, to address the issue of preference discrepancy, the uncertainty set is required lying around the true user preference distribution. A reasonable approach to construct the uncertainty set is to introduce the \emph{nominal distribution} \cite{DROsurvey,hu2013kullback, OODsurvey} which contains the prior knowledge regarding the true user preference, as shown in Fig. \ref{intorduction-b}. 
Here, we adopt the system exposure distribution as the nominal distribution. 
The insight behind this choice lies in the following points: 1) system exposure can be viewed as the user preference from the recommender perspective and lies around the true user preference; and 2) system exposure contains prior knowledge about both the recommender exposure mechanism and the true user preference distribution.
Besides, since exposure bias occurs when a recommender system exposes items that are deemed by the system as preferred by a user but are not actually liked by the user, there is an overestimation of user preference on biased samples.
Finally, due to the unavailability of unbiased sequential recommendation datasets~\cite{Khalil2022cloze}, how to evaluate the debiasing effect on the offline biased test set is also a problem.  To this end, this paper aims to address the following research challenges:

\noindent\textbf{Research challenges.}
i). How to model the system exposure distribution as the nominal distribution of DRO? 
To address this challenge, we pre-train an exposure simulator by modeling the previous system behavior data (i.e., exposed items).
The task of simulating system exposure is challenging since the exposure mechanism typically involves the combination of multiple recommenders.
Besides, service providers frequently prioritize the recommendation of popular (hot) items, further complicating the process of simulating. To this end, we model the exposure distribution by the mixture of multiple recommender models
\footnote{In this paper, we utilize three recommender models to model the sequential exposure distribution. We leave the investigation of more advanced exposure simulator as one of our future works.}.
    Remarkably, this paper represents the first attempt, to the best of our knowledge, to model the exposure distribution explicitly in the context of sequential recommendation.

ii). How to eliminate the overestimation of user preference on biased samples?  For this challenge, we impose a penalty on items that the system deems as highly preferred by the user. This penalty can reduce the influence of such items and enable the system to make more accurate predictions of user preference. To more accurately identify samples with exposure bias, we penalize items according to their exposure distribution, applying a larger penalty to items with high exposure probabilities.

iii). How to evaluate the debiasing effect on the offline biased test set? For this challenge, we design a self-normalized inverse propensity score (SNIPS) \cite{snips2015,ips_rec, Yang2018unbiasevaluatuion} evaluator to evaluate the debiasing effect on the biased test set. The idea of SNIPS is to down-weight frequently exposed samples and up-weight the rare ones. 
Here we treat the exposure probability of each item as the corresponding propensity score. 
Nevertheless, using the prediction of the pre-trained exposure simulator as exposure probability may cause potential data leakage for evaluation. 
To solve this issue, we construct an extra exposure simulator based on another unseen part of the recommender exposure to generate the propensity score. 

\noindent\textbf{Experiments.} Extensive experiments and analyses on two real-world datasets demonstrate the effectiveness of the proposed framework. We hope this work can raise more community concern regarding improving the recommenders from the system behavior perspective other than just focusing on debiasing from the user behavior perspective. 

\noindent\textbf{Contributions.} To summarize, the main contributions are: 
\begin{itemize}[leftmargin=*,nosep]
\item We propose to use Distributionally Robust Optimization to conduct exposure debiasing in sequential recommendation.
\item  In contrast to only considering debiasing from the user perspective, we incorporate the system behavior and alleviate the exposure bias from perspectives of both users and systems.
\item We design a debiased self-normalized inverse propensity score (SNIPS)  evaluator for evaluating the debiasing sequential recommendation performance in the biased test set.
\end{itemize}
\section{Related work}
\label{sec:related work}
In this section, we provide a literature review regarding sequential recommendation and exposure debiasing.
\subsection{Sequential Recommendation}
Recently, sequential recommendation (SR) has become a hot research topic. Compared with the other task in recommendation \cite{koren2009mf,ning2011slim}, the SR focuses on capturing the item's chronological correlations. Many sequential recommender methods have been proposed, such as methods based on the Markov Chain~\citep{DBLP:conf/www/RendleFS10,DBLP:conf/recsys/HeFWM16} and factorization methods \cite{gmf}. Over the past few years, plenty of deep learning-based SR models have emerged, including Recurrent Neural Networks (RNNs)-based methods \cite{DBLP:journals/corr/HidasiKBT15,DBLP:conf/recsys/HidasiQKT16}, Convolutional Neural Networks (CNNs)-based methods \cite{nextitnet}, and attention-based methods \cite{NARM, SASRec}. 
Besides, Graph Neural Networks (GNNs) also demonstrated their capability to represent user-item high-order interactions in the SR. Plenty of graph-based SR models have emerged, such as SR-GNN \cite{wu19srgnn}, UGrec \cite{wang19ugrec}, and SURGE \cite{chang2021SURGE}. 
Casual inference \cite{wang2022lantentconfounders}, reinforcement learning \cite{xin2020self}, and self-supervised learning \cite{CIKM2020-S3Rec} also demonstrated the effectiveness to enhance SR.

In this paper, instead of designing a new recommender model, we focus on mitigating the impact of exposure bias in the SR. To this end, we propose a model-agnostic debiasing framework and adopt two representative models, SASRec and GRU4Rec, as backbone sequential recommendation models for our experiments. 

\subsection{Exposure Debiasing in RSs}
\label{subsection:exposure_debias}
Exposure bias, also known as "previous model bias" ~\cite{DBLP:conf/sigir/LiuCDHP020}, occurs when user interactions are dependent upon the partial items exposed to the user.
As it is not clear whether a lack of interaction between a user and an item is due to genuine disinterest or simply a consequence of no exposure, a non-exposed item may actually be favored by the user but unable to be interacted with.
Exposure bias would mislead both the model's training and evaluation \cite{biassurvey}.
The common approach for debiasing recommender system evaluation is incorporating the well-known \emph{Inverse Propensity Scoring} (IPS) \cite{ips_rec} into the ranking evaluation metrics. More specifically, the key idea of IPS-based evaluation is that items are down-weighted by their propensity in the evaluation metrics. For instance, \cite{Yang2018unbiasevaluatuion} proposes an IPS-based unbiased evaluator to down-weight the commonly observed interactions, while up-weighting the rare ones. 
Furthermore, various strategies are being used to mitigate exposure bias during the training phase. In early works, researchers aim to integrate a measure of confidence into the unobserved interactions instead of considering them as irrelevant. For example, \cite{DBLP:conf/kdd/PanS09} utilizes user activity (e.g., the number of interacted items) to weigh the negative interactions. 
\cite{DBLP:conf/wsdm/SaitoYNSN20} argues that early methods can not address exposure bias entirely and designed the IPS-based unbiased estimator in the training phase to perform debiasing.
\cite{10.1145/3306618.3314288} proposes an exposure-based propensity matrix factorization framework to counteract the exposure bias. 

Recently, there have been some works \cite{wang2022lantentconfounders, Khalil2022cloze,DBLP:conf/cikm/Xu0CDW22} to remedy bias in SR. Since the interaction distribution in SR evolves over time, it is reasonable to assume that exposure propensities also vary over time \cite{Khalil2022cloze}. Thus, the traditional IPS  does not applicable to SR because it fails to account for the temporal nature of the problem. To this end, \cite{wang2022lantentconfounders} uses GRU \cite{GRU} to estimate temporal IPS, and derived an IPS-based loss function via potential latent confounders in RS. 
\cite{DBLP:conf/cikm/Xu0CDW22} estimates the propensity scores with two GRUs from the views of items and users in SR.
\cite{Khalil2022cloze} proposes a temporal IPS-based framework for debiasing the Cloze task \cite{sun2019bert4rec} of SR. 

While existing methods have explored alleviating the exposure bias problem, they only rely on user interaction data to estimate the exposure policies and neglect the system perspective. 
In this work, we propose to simulate the exposure mechanism of the real-world recommendation system by considering both interacted data from the user perspective and exposure data from the system perspective. 
Besides, compared with the high variance IPS, the adopted DRO framework provides more robust and stable model training. 
To the best of our knowledge, our work is the first attempt to utilize DRO for exposure debiasing in sequential recommendation.

\section{Notations and Problem Formulation}
\label{sec:preliminay}
\header{Sequential recommendation.}
Let $\mathcal{U}$ and $\mathcal{I}$ denote the user set and the item set, respectively, where $u \in \mathcal{U}$ denotes a user and  $v \in \mathcal{I}$ denotes an item.
We use $S^u$=($v^u_1$, $v^u_2$, ..., $v^u_{t}$) to denote the sequence of user $u$'s previous interactions (e.g., views, clicks, or purchases) in the sequence, where $v^u_{j}$ is the $j$-th item user $u$ has interacted with, $t$ is the length of the interaction sequence. 
 SR aims to predict the next item that the user is likely to interact with at the $(t+1)$-th step, which can be formulated as the estimation of $P(v_{t+1}|S^u)$, where $P$ denotes the user’s preference score for the item $v$ in the step $(t + 1)$.
The prediction error is measured by the prevalent Binary Cross Entropy (BCE) \cite{SASRec, CIKM2020-S3Rec} loss:
\begin{equation} 
    \label{eq:rec}
    \begin{small}
    \mathcal{L}_{\text {rec}}\!=\!-\!\sum_{\!u \!\in \!\mathcal{U}} \!\sum_{j \!\in [\!2,\!...,t]} \![\log \sigma\!\left(P\!\left(\!v_{j}^{+}| S^u_{<\!j}\right)\right)+\!\log \!\sigma\!\left(1-\!P\!\left(v_{j}^{-}| S^u_{<\!j}\right)\right)],
    \end{small}
\end{equation}
where $v_{j}^{+}$ and $v_{j}^{-}$ denote the positive sample and negative sample at the $j$-th step, respectively. The $S^u_{<j}$ denotes the sequence of user $u$ before $j$-th step. $\sigma$ denotes the Sigmoid activation function.

\header{System exposure data.} Compared with the sparse user interaction data, the large volume of exposure data generated by the recommender receives relatively less research attention. Given an $n$-length exposure item sequence $E^u=(e^u_1, e^u_2, ..., e^u_{n})$, where $e^u_{j}$ represents the $j$-th item exposed to user $u$ and $n$ is the length of exposure sequence.  For a user $u$, the user interaction data $\{S^u\}$ is a subset of $\{E^u\}$ since the exposure data include both interacted and non-interacted but exposed items. The exposed items in system exposure data are selected by the recommender and thus can be regarded as a kind of system behavior data. In this paper, we construct an exposure simulator by modeling the system behavior, aiming to simulate the system exposure distribution.

\header{Task formulation.} The task of this paper is given the user-item interaction sequence $S^u$ and the system exposure sequence $E^u$, eliminating the impact of exposure bias and generating the recommendation list for user $u$.

\section{methodology}
\label{sec:methodology}
\begin{figure*}[t]
  \centering
  \includegraphics[width=0.7\linewidth]{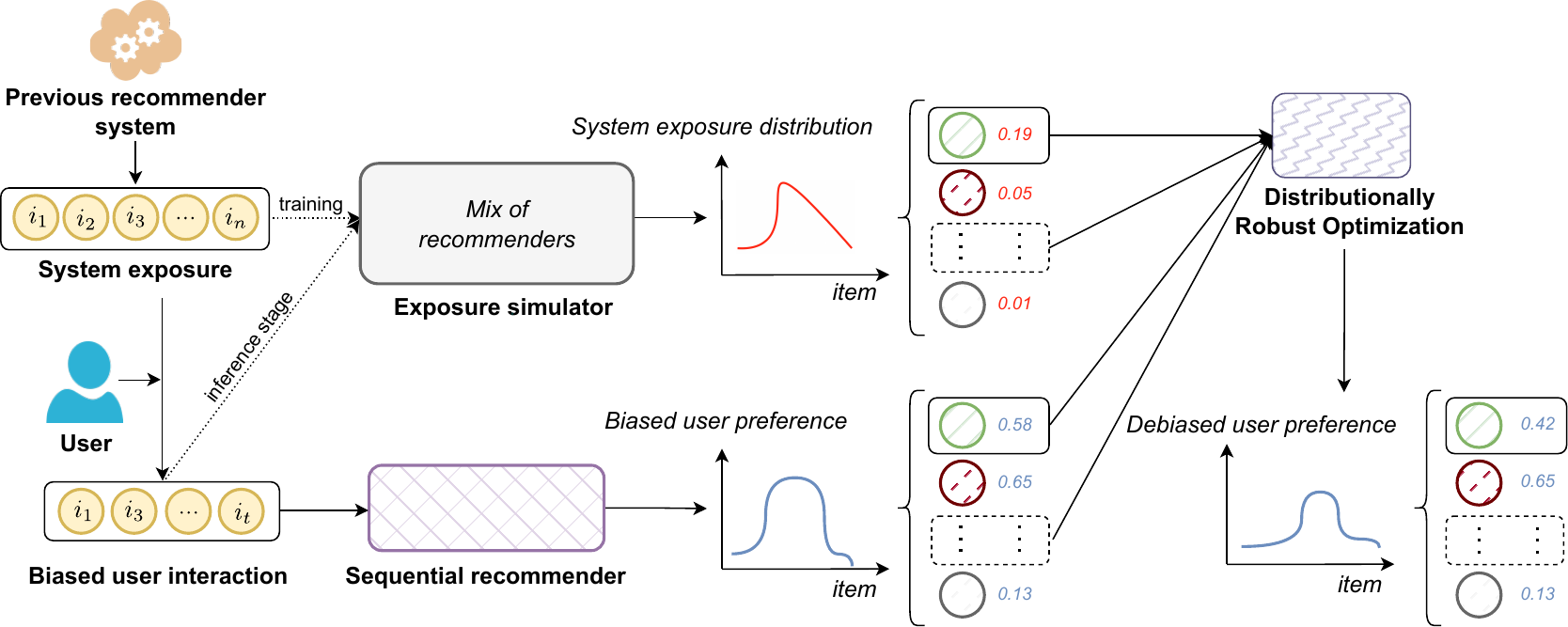}
  \caption{Framework overview. For the user side, users observe exposed items from previous recommenders, generating biased interactions.
  For the system side, exposure data is used to train the exposure simulator and obtain the exposure distribution. 
  During inference, the biased user preference and the system exposure distribution are fed to the DRO block for debiasing.}
  \label{fig:framework}
\end{figure*}

In this section, we present the detail of our proposed debiasing framework.  
Fig. \ref{fig:framework} provides the overview of our methods, which includes both the user and system perspectives. From the \emph{user} perspective, users observe the exposed items and interact with specific parts of them, generating biased user-item interaction data. 
Therefore, the sequential recommender trained on such biased data would lead to biased user preference. 
From the \emph{system} perspective, we model the distribution of exposure data by pre-training an exposure simulator to mimic the real-world exposure mechanism. The exposure simulator aims to provide a system behavior distribution (i.e., the exposure probability of each item). We then utilize the system exposure distribution as the nominal distribution of Distributionally Robust Optimization (DRO) to infer the debiased user preference and generate sequential recommendation.
It is important to note that the sequential recommender is jointly trained by user interaction data and the DRO method, the exposure model is only pre-trained by system exposure data and its parameters are fixed in the inference stage. 

\label{subsec:framework}

\subsection{Distributionally Robust Optimization}
\label{subsection:DRO}
Distributionally Robust Optimization (DRO) \cite{DROsurvey,DBLP:journals/corr/abs-1911-08731,slowik2022distributionally} has been widely adopted as a solution to the out-of-distribution (OOD) \cite{OODsurvey} generalization problem ~\cite{wen2022distributionally}. The key idea of DRO is to optimize the worst-case error over an \emph{uncertainty set} to safeguard the model against potential distributional shifts within the uncertainty set \cite{OODsurvey}. The SR model is typically trained on biased user-item interactions. It means that the preference generated by the biased recommender may not align with the true user preference. 
In this paper, we employ DRO to alleviate the impact of the distribution discrepancy between biased user preference and unknown true user preference in SR.
The problem of distribution shifting is even more pronounced in the context of SR, where users' preference dynamically change.

DRO replaces the expected risk under the empirical distribution (i.e., biased interaction distribution) with the worst expected risk over a set of uncertain distributions:
\begin{equation}
    \label{eq:dro}
    \mathcal{L}_{DRO}(\theta)=\max _{\hat{\mathcal{D}} \in \mathcal{Q}} \mathbb{E}_{(S^u, v) \sim \hat{\mathcal{D}}} \ell(S^u, v; \theta),
\end{equation}
where $\theta$ refers to the parameters to be optimized and $\ell(S^u,v;\theta)$ is the surrogate loss and will be detailed in section \ref {subsection:Over-estimation}.
The DRO loss function $\mathcal{L}_{DRO}$ takes the maximum over all distributions contained in a pre-defined \emph{uncertainty set} $\mathcal{Q}$.
$\mathcal{L}_{DRO}$ can be interpreted as an upper bound estimation of the average loss on the true user preference distribution if the distribution lies in $\mathcal{Q}$.
A viable strategy is to form the uncertainty set $\mathcal{Q}$ by stipulating that the distribution is within a certain proximity to the nominal distribution $q_0$, which is an approximation of the true user preference distribution and encapsulates important information about user preference.
To this end, we introduce the distance metric $D$ (i.e., the Kullback–Leibler Divergence) to the DRO loss function in Eq.(\ref{eq:dro}), so $\mathcal{Q} = \left\{q:  D\left(q \| q_{0}\right) \leq \eta\right\}$, where $\eta$ is the robust radius to control the size of $\mathcal{Q}$. 
We can then reformulate the DRO loss function as:
\begin{equation}
    \mathcal{L}_{DRO}^{'}(\theta)=\max _{D\left(q \| q_{0}\right) \leq \eta} \mathbb{E}_{(S^u, v) \sim q} \ell(S^u, v; \theta),
    \label{eq:dro1}
\end{equation}
To address the issue of preference discrepancy, the nominal distribution is required to lie around the true preference distribution. In this paper, we adopt system exposure distribution as the nominal distribution.
The rationale behind this choice lies in the following points: 1) system exposure can be viewed as the user preference from the system perspective and lying around the true user preference, and 2) system exposure distribution contains prior knowledge about the recommender exposure mechanism that can be leveraged to perform exposure debiasing. Specifically, the exposure probability for item $v$ is denoted as $q_0(S^u, v)$.

\subsection{Modeling Exposure Distribution}
\label{subsec:model_exposure}
The system exposure distribution can be considered as the system behavior patterns and reflects the system's understanding of user preference. 
Existing exposure debiasing approaches in~\cite{DBLP:conf/wsdm/SaitoYNSN20,10.1145/3306618.3314288, Yang2018unbiasevaluatuion} rely solely on observed interaction data to estimate the unobserved exposure mechanism in implicit feedback recommendation from the user perspective. However, we argue that these approaches suffer from sub-optimal performance as they did not introduce the large volume of exposure data from the system perspective.
In contrast to existing methods, we simulate the exposure mechanism of the real-world recommender system by considering both interacted data and exposed but non-interacted data. 

Simulating the real-world exposure mechanism is a daunting task due to two key factors: 1) the real-world exposure mechanism often entails a complex interplay of multiple recommenders, making it challenging to accurately model the precise exposure of individual items; 2) recommendation service providers tend to prioritize popular items to increase the traffic, further complicating the simulation process. To this end, we develop an exposure simulator that consists the mixture of multiple recommenders.  The simulator comprises a Transformer-based~\cite{Transformer} recommendation model (SASRec \cite{SASRec}), a GRU-based~\cite{GRU} recommendation model (GRU4Rec \cite{DBLP:conf/recsys/HidasiQKT16}), and a popularity-based recommender that boosts popular items.
The three recommenders in the simulator can be replaced by other recommendation methods. We leave the investigation of a more advanced exposure mechanism as one of our future works. 
\subsubsection{Training of the exposure simulator.}
For the training stage, it is worth noting that the training of the exposure simulator differs from that of sequential recommendation, primarily because:
\begin{itemize}[leftmargin=*,nosep]
    \item The input of the exposure simulator in the training stage is the \emph{system} exposure data while the input for sequential recommendation is the \emph{user} interactions.
    \item The output of the exposure simulator in the training stage is the \emph{system} exposure distribution while the output of sequential recommendation is the predicted \emph{user} interests.
\end{itemize}
In the training stage, we use the system exposure data to train the SASRec model (denoted as Expo-SASRec)  and the GRU4Rec model (denoted as Expo-GRU4Rec), separately. Then the parameters in the pre-trained Expo-SASRec and Expo-GRU4Rec remain fixed in the following stages to provide stable exposure simulation. More specifically, the Expo-SASRec model is based on a Transformer decoder \cite{SASRec}. Expo-GRU4Rec is an RNN-based sequential recommender that uses GRU to encode users' interaction sequences, consisting of an embedding layer, recurrent layers, and fully connected layers. 

\subsubsection{Exposure distribution inference.}
It is important to note that, in the inference stage, the
input to the exposure simulator is the user interaction data rather than system exposure data. The motivation is that during inference, the system exposure distribution should depend on which items the user have interacted with. 
The goal of such design is to gain insights into user sequential interests from a system perspective. 

Given a user-item interaction sequence $S^u$=($v^u_1$, $v^u_2$, ..., $v^u_{t}$),
the hidden state of the Expo-SASRec encoder in the last timestamp serves as the latent representation of the input sequence: $\mathbf{F}^u_{t,1} = \text{Expo-SASRec}(S^u)$.
$\mathbf{F}^u_{t,1}$ can be viewed as the user's potential preference representation from the system's (Expo-SASRec) perspective at the $t$-th step. 
Similarly, we also leverage Expo-GRU4Rec to encode user interaction data, The exposure representation through Expo-GRU4Rec is denoted as: $\mathbf{F}^u_{t,2} = \text{Expo-GRU4Rec}(S^u)$. 
Additionally, we introduce a popularity-based model to expose more popular items. We define the popularity score of item $i$ as
$f_i=\frac{s_i}{\max_{j \in \mathcal{I}} s_j}$ where $s_i$ denotes that item $i$ appears $s_i$ times in the exposure data. 
Finally, we calculate the recommender exposure score for the item $v$ at the step ($t$+1) as follows:
\begin{equation}
    \setlength{\abovedisplayskip}{3pt}
    \setlength{\belowdisplayskip}{3pt}
    \begin{split}
    \label{eq:predict}
    \mu_0(S^u, v_{t+1} = v) &= softmax(\mathbf{e}_v^{\top} \cdot \mathbf{F}^u_{t,1}) + softmax(\mathbf{e}_v^{\top} \cdot \mathbf{F}^u_{t,2}) \\
    &+ \beta \ softmax(f_v),
    \end{split}
 \end{equation}
where $\mathbf{e}_v$ is the embedding of item $v$ and $\beta$ is a hyperparameter.

We calculate the exposure distribution as follows:
\begin{equation}
    \setlength{\abovedisplayskip}{3pt}
    \setlength{\belowdisplayskip}{3pt}
    \label{eq:exposure_probability}
    q_0(S^u,v) = \frac{\mu_0(S^u,v)}{\sum_{i \in \mathcal{I}} \mu_0(S^u,i)}.
\end{equation}
A high exposure probability $q_0(S^u, v)$ indicates a significant level of exposure. In summary, the proposed method of modeling exposure distribution offers the following advantages:
\begin{itemize}[leftmargin=*,nosep]
    \item 
    Instead of relying on statistical methods to estimate exposure policy, we choose to explicitly model the exposure distribution through exposure data. Our methods provide more adaptive modeling of exposure distribution.
    \item 
    The simulator is based on the mixture of sequential models and item popularity, which is more effective to capture both sequential dynamics and popularity effects in the exposure distribution.
\end{itemize}

\subsection{Mitigating Preference Overestimation}
\label{subsection:Over-estimation}
From a system perspective, exposure bias occurs when a recommender system exposes items that are deemed by the system as preferred by a user but are not actually preferred by the user, resulting in the failure to expose items that are actually preferred by the user. Therefore previous system exposure can not accurately obtain current user preference, i.e., the system assigns a higher score to an item than the user's true preference. In other words, the system exposure distribution may overestimate the scores of biased samples that are not aligned with the user's true preference. To solve this issue, we optimize the recommender to mitigate the risk of preference overestimation and pursue conservative recommendation under the DRO framework.

Note that minimizing the DRO loss is an intractable min-max problem. To address this limitation, we utilize the Kullback-Leibler (KL) Divergence as the metric for measuring distance denoted as $D$ in Eq.(\ref{eq:dro1}). Through this approach, we can derive a closed-form expression for Eq.(\ref{eq:dro1}) 
as a single-layer optimization problem as:
\begin{equation}
    \setlength{\abovedisplayskip}{3pt}
    \setlength{\belowdisplayskip}{3pt}
    \label{eq:dro2}
    \mathcal{L}_{DRO}^{''}(\theta)=\log \mathbb{E}_{(S^u, v) \sim q_{0}} e^{\ell(S^u, v; \theta )} .
\end{equation}
For detailed proof of the reformulation, please refer to Appendix \ref{subsec:appendix_proof_dro}.
The surrogate loss $\ell(S^u,v;\theta)$ is also called the risk function of DRO.
The risk can be categorized into the following two cases:
1) if ($S^u$, $v$) is a positive sample, it means that the user is possibly to like the item, and thus the risk is high if the sample receives a lower prediction score; 2) if ($S^u$, $v$) is a negative sample, a higher prediction score denotes a higher risk of preference overestimation. 

To this end, we define the surrogate loss $\ell(S^u,v;\theta)$ as follows:
\begin{equation}
\label{eq:risk_fun}
\ell(S^u,v;\theta) =\left\{\begin{array}{ll}
1-y(S^u,v; \theta), & \text{if }(S^u, v) \text { is a positive} \\
y(S^u,v; \theta), & \text {if }(S^u, v) \text { is a negative},
\end{array}\right.
\end{equation}
where $y(S^u,v; \theta)$ denotes the prediction score of the $(S^u, v )$ pair in the DRO block.
In our setting, the number of negative samples is $|\mathcal{I}|-1$, which encompasses all items apart from positive samples. 
We can see optimizing Eq.(\ref{eq:dro2}) equals to sample $(S^u, v)$ from the nominal exposure distribution $q_0$, then if the sampled $(S^u, v)$ pair is a negative sample, we would decrease its predicted score. Such the operation can help the model to mitigate the risk of preference overestimation.

To conclude, optimizing Eq.(\ref{eq:dro2}) can enhance the robustness of the recommender under the dynamic exposure bias, but pursuing robustness without considering recommendation accuracy may significantly compromise the system's performance. 
To this end, we jointly consider the recommender robustness in Eq.(\ref{eq:dro2}) and accuracy in Eq.(\ref{eq:rec}), and the final optimization objective can be defined as:
\begin{equation}
\begin{split}
   \label{eq:training_loss} \mathcal{L}_{joint}&=\mathcal{L}_{rec}+a\mathcal{L}_{DRO}^{''} \\
   &= \mathcal{L}_{rec} + a\sum_{u \in \mathcal{U}} \sum_{j \in [2,...,t]} 
   \log \mathbb{E}_{(S^u_{<j}, v) \sim q_{0}} e^{\ell(S^u_{<j}, v; \theta )},
\end{split}
\end{equation}
where $a$ is a hyperparameter, $v^+$ and $v^-$ denote positive and negative samples, respectively. Note that the $P$ in $\mathcal{L}_{rec}$ and $y$ in Eq.(\ref{eq:risk_fun}) are both predicted preference scores. $P$ is calculated from the target sequential recommender, while $y$ is calculated from a model which has a separate final prediction layer and shares all other parameters with the target sequential recommender.
The joint optimization objective can be viewed as two tasks: recommendation and exposure debiasing. 
By optimizing the recommender using the joint loss function, we can approximate the user's true preference meanwhile mitigate the exposure bias. It is worth noting that the $\mathcal{L}_{joint}$ can serve as a surrogate for any sequential recommendation model, making this approach model-agnostic.

\subsection{The Debiased Evaluator}
\label{evaluator}
To evaluate the debiasing effect, we need reliable debiased test set~\cite{DBLP:conf/wsdm/SaitoYNSN20,Khalil2022cloze}. However, most existing benchmark test sets are also affected by the exposure bias \cite{Khalil2022cloze}.
An ideal case is to access complete observations, i.e., user $u$ has the chance to interact with the entire set of items (i.e., $\mathcal{I}$). Therefore, the ideal recommendation evaluator calculates the following reward $R_{\text{ideal}}(\hat{Z})$ for the predicted item ranking $\hat{Z}$:
\begin{equation}
    \setlength{\abovedisplayskip}{3pt}
    \setlength{\belowdisplayskip}{3pt}
    \label{ideal}
    R_{\text{ideal}}(\hat{Z})=\frac{1}{|\mathcal{U}|} \sum_{u \in \mathcal{U}} \frac{1}{\left|\mathcal{I}\right|} \sum_{i \in \mathcal{I}} c\left(\hat{Z}_{u, i}\right),
\end{equation}
where $i \in \mathcal{I}$ is an item, and $\hat{Z}_{u, \, i}$ is the predicted ranking of item $i$ (among all the items in $\mathcal{I}$) for user $u$. $c(\cdot)$ represents any of the top-$K$ scoring metrics, such as Recall@$K$ or NDCG@$K$.

However, users have no chance to interact with the entire set of items, leading to biased observations. To achieve a debiased evaluation from biased observations, we employ a debiased self-normalized inverse propensity score (SNIPS). The idea is to down-weight frequently exposed interactions and up-weight the rare ones in the test set.
The SNIPS evaluator is defined as follows:
\begin{equation}
    \setlength{\abovedisplayskip}{3pt}   
    \setlength{\belowdisplayskip}{3pt}   
\begin{split}
    \label{eq:SNIPS}
    \hat{R}_{\text{SNIPS}}(\hat{Z}|\rho) &= \frac{1}{\sum_{u \in \mathcal{U}}{\frac{1}{\rho^{k}_{v}}}} \sum_{u \in \mathcal{U}}\frac{c(\hat{Z}_{u, \, v)}}{\rho^{k}_{v}} \\
    &=\frac{1}{\sum_{u \in \mathcal{U}}{\frac{1}{\rho^{k}_{v}}}} \sum_{u \in \mathcal{U}} \sum_{i \in \mathcal{I}} \frac{c(\hat{Z}_{u, \, i)}}{\rho^{k}_{i}} \cdot O_{u,i},
\end{split}
\end{equation}
 where $\rho$ is the propensity score of the SNIPS evaluator, $v$ is the target item of each user $u$, and $O_{u,i}$ is an indicator function to indicate whether the interaction $(u, i)$ is observed ($O_{u,i} =1$ if $(u, i)$ is observed, and $O_{u,i} =0$ otherwise). 
 Moreover, we propose the inclusion of a hyperparameter $k\in [0,1]$, which enables control over the level of the propensity score. 
We prove that given a propensity assignment $\rho$,  $\hat{R}_{\text{SNIPS}}(\hat{Z}|\rho)$ is a debiased estimator (More details of proof can be found in Appendix \ref{subsec:appendix_proof_snips}.):
\begin{equation} 
    \begin{split}
    \setlength{\abovedisplayskip}{2.5pt}
    \setlength{\belowdisplayskip}{2.5pt}
    \label{eq:E_SNIPS}
    \mathbb{E}_O\![\hat{R}_{\text{SNIPS}}\!(\!\hat{Z}|\rho)\!] 
    &=\! \frac{|\mathcal{U}| \cdot |\mathcal{I}|}{\!\sum_{u \in \mathcal{U}}{\frac{1}{\rho^{k}_{v}}}} \!\sum_{u \in \mathcal{U}} \!\sum_{i \in \mathcal{I}} \frac{1}{\rho^{1-k}_{i}} \!\cdot \mathbb{E}_O[R_{\text{ideal}}(\hat{Z})]. 
    \end{split}
\end{equation}
The term $\frac{|\mathcal{U}| \cdot |\mathcal{I}|}{\sum_{u \in \mathcal{U}}{\frac{1}{\rho^{k}_{v}}}}$ is a coefficient which not affect the unbiasedness of the evaluator. The evaluator is a conventional biased one when the parameter $k$ is set to 0. When the parameter $k$ is set to 1, the evaluator is completely unbiased,  but this comes at the cost of increased variance of the propensity score, manifesting as noisier or less precise estimates.
In our experiments, $k$ is set to 0.1.

The key challenge in computing $\hat{R}_{SNIPS}(\hat{Z})$ is to predict the propensity score $\rho$. Here we treat the exposure probability of each item as the corresponding propensity score since we aim to evaluate the exposure debiasing performance using the SNIPS evaluator. Nevertheless, using the prediction of the pre-trained exposure model (as detailed in section \ref{subsec:model_exposure}) as exposure probability may cause training data leakage. More concretely, if the exposure model is trained by all the system exposure data, the testing user interaction data would be leaked because all the test data is contained in exposure data. To solve this issue, we construct an extra evaluation exposure simulator based on another unseen part of system exposure to predict the exposure probability in the testing phase.

\section{Experiments}
\label{sec:experiments}
In this section, we conduct experiments to demonstrate the effectiveness of the proposed debiasing framework. We aim to answer the following research questions:
\begin{enumerate}[leftmargin=*, label=RQ\arabic*]
\item How does our proposed debiasing framework perform compared to existing debiasing methods?

\item How does the construction of the exposure simulator affect debiasing performance?

\item How does the penalty term in the DRO loss affect the debiasing performance?
\end{enumerate}
\subsection{Dataset Description}
\label{subsection:datasset}
We choose two real-world datasets: ZhihuRec\footnote{\url{https://github.com/THUIR/ZhihuRec-Dataset}}~\citep{hao2021largescale} and Tenrec\footnote{\url{https://github.com/yuangh-x/2022-NIPS-Tenrec}}~\citep{tenrec} to conduct the experiments. Both datasets contain user interaction data (e.g., clicks) and system exposure data (e.g., impressions). All items in the sequence are arranged chronologically based on their timestamp. More details can be found in Appendix \ref{sec:appendix_dataset}. 
\subsection{Evaluation Protocols}
\label{subsec:Evaluation}
We employ the debiased self-normalized inverse propensity score (SNIPS) \cite{snips2015,ips_rec, Yang2018unbiasevaluatuion} evaluator as detailed in section \ref{evaluator} to gauge the effectiveness of debiasing.
We adopt Recall and NDCG as the $c(\cdot)$ function metrics to evaluate the ranking performance. 
We report the results with varying values of $K$ $\in$ \{5, 10, 20\} for both metrics.
To prevent information leakage, it is crucial to avoid any overlap in the system exposure data used for constructing the exposure simulator and the evaluation simulator. 
We split the system exposure data into two parts: with 70\% being used to construct the exposure simulator and the remaining 30\% for constructing the evaluation simulator. 

We adopt cross-validation to evaluate the performance of the models (including the target recommender, exposure simulator, and evaluation simulator). 
For all models, we choose  80\% users' data for training,  10\% users' data for validation, and the rest 10\% users' data for test.
Such user-based data split can effectively avoid potential information leakage.
 
\subsection{Implementation Detail}
\header{Backbone models.} The proposed method is model-agnostic, we adopt the following representative sequential recommender models as backbone models for our experiments:
\begin{itemize}[leftmargin=*,nosep]
    \item GRU4Rec \cite{DBLP:conf/recsys/HidasiQKT16} is an RNN-based sequential recommender, which leverages GRU \cite{GRU} to encode users’ interaction sequences.
    \item SASRec \cite{SASRec} leverages a Transformer decoder to generate sequential recommendation.
\end{itemize}
See Appendix \ref{appedix:reproducibility} for hyperparameter settings of backbone models.

\header{Baselines.}
We compare the exposure debiasing performance of our method with the following state-of-the-art debiasing methods:
\begin{itemize}[leftmargin=*,nosep]
    \item IPS \cite{ips_rec, CausalInferenceSurvey2023}: IPS was first developed by \cite{ips_rec} to eliminate popularity bias by re-weighing each interaction according to propensity score.
    \item IPS-C \cite{DBLP:conf/www/ZhengGLHLJ21,DBLP:journals/jmlr/BottouPCCCPRSS13}: 
IPS-C adds a max-clipping operation on the IPS value to reduce the variance of IPS.
    \item RelMF \cite{DBLP:conf/wsdm/SaitoYNSN20}: The method aims to use an effective unbiased estimator to correct the matching score between items and users.
\end{itemize}
Note that the propensity scores in the original paper are solely estimated by the user interaction data and fail to account for the temporal nature of the problem. To solve this issue and adapt these methods to exposure debiasing in SR, we define propensity scores $p(v|S^u) = q_0(S^u,v)$, where $q_0$ is the system exposure distribution as detailed in subsection \ref{subsec:model_exposure}. Following \cite{DBLP:conf/www/ZhengGLHLJ21}, the clipping severity in IPS-C is set to the median exposure probability of all the items. Besides, we also include the normal training of GRU4Rec and SASRec without applying any debiasing methods for comparison purposes. 

\subsection{Overall Debiasing Performance (RQ1)}
\label{sec:mainresults}
\begin{table*}[ht]
    \centering
    \setlength{\abovecaptionskip}{3pt}
    \begin{threeparttable}
    \caption{Comaprison of debiased recommendation performance. Boldface denotes the highest score. ``None'' denotes the normal training. ``DRO'' denotes our method.}
    \renewcommand\arraystretch{0.9}
    \label{table:rq1}
    \begin{tabular}{ccc|cc|cc|cc}
    \toprule
Datasets&Backbone&Method&Recall@5&NDCG@5&Recall@10&NDCG@10&Recall@20&NDCG@20\cr
    \midrule
    \multirow{10}{*}{ZhihuRec}
    &\multirow{5}{*}{GRU4Rec}
    &None  & 0.0094& 0.0072& 0.0114& 0.0079& 0.0231& 0.0108\cr
    &&IPS  & 0.0083& 0.0070& 0.0137& 0.0087& \textbf{0.0244}& \textbf{0.0113}\cr
    &&IPS-C& 0.0083& 0.0057& 0.0114& 0.0067& 0.0189& 0.0085\cr
    &&RelMF& 0.0032& 0.0021& 0.0106& 0.0046& 0.0201& 0.0069\cr
    &&DRO  & \textbf{0.0103} & \textbf{0.0076}& \textbf{0.0146}& \textbf{0.0090}& 0.0200& 0.0104\cr
    \cmidrule(lr){2-9}
    
    &\multirow{5}{*}{SASRec}
    &None  & 0.0119& 0.0089& 0.0162& 0.0104& 0.0241& 0.0123\cr
    &&IPS  & 0.0109& 0.0079& 0.0210& 0.0112& 0.0450& 0.0171\cr
    &&IPS-C& 0.0151& 0.0097& 0.0174& 0.0104& 0.0261& 0.0126\cr
    &&RelMF& 0.0095& 0.0074& 0.0138& 0.0088& 0.0257& 0.0116\cr
    &&DRO  & \textbf{0.0220}& \textbf{0.0131}& \textbf{0.0322}& \textbf{0.0162}& \textbf{0.0498}& \textbf{0.0205}\cr
    \hline
    \multirow{10}{*}{Tenrec}
    &\multirow{5}{*}{GRU4Rec}
    &None  & 0.0335& 0.0218& 0.0524& 0.0280& 0.0735& 0.0333\cr
    &&IPS  & 0.0320& 0.0210& 0.0513& 0.0273& 0.0706& 0.0322\cr
    &&IPS-C& 0.0381& 0.0244& 0.0533& 0.0293& \textbf{0.0802}& \textbf{0.0361}\cr
    &&RelMF& \textbf{0.0382}& \textbf{0.0251}& 0.0523& \textbf{0.0296}& 0.0767& 0.0357\cr
    &&DRO  & 0.0336& 0.0225& \textbf{0.0536}& 0.0290& 0.0748& 0.0343\cr
    \cmidrule(lr){2-9}
    
    &\multirow{5}{*}{SASRec}
    &None  & 0.0337& 0.0216& 0.0638& 0.0313& 0.1044& 0.0414\cr
    &&IPS  & 0.0365& 0.0223& 0.0647& 0.0314& 0.1041& 0.0414\cr
    &&IPS-C& 0.0410& 0.0266& 0.0735& 0.0370& 0.1143& 0.0473\cr
    &&RelMF& 0.0363& 0.0238& 0.0499& 0.0282& 0.0752& 0.0346\cr
    &&DRO  & \textbf{0.0517}& \textbf{0.0335}& \textbf{0.0902}& \textbf{0.0457}& \textbf{0.1439}& \textbf{0.0592}\cr
    \bottomrule
    \end{tabular}
    \end{threeparttable}
\end{table*}

Table \ref{table:rq1} shows the overall debiasing performance.
We observe that the DRO method outperforms both vanilla backbone models and debiasing baselines on two datasets in most cases. These findings suggest that DRO is able to effectively mitigate the risk of overestimating biased samples and conduct exposure debiasing. 
Note that the data is partitioned by the user, with no overlap between the users in the training and test sets. In other words, user preference in the training and test data could have shifted in the sequential recommendation scenario. If the recommender overly relies on the preference learned from training users, it may raise the overestimation issue of the test users' preference and result in sub-optimal performance.
Compared with other methods, the proposed DRO framework compels the model to be less reliant on the training data distribution and to make more conservative predictions through the surrogate loss, resulting in better recommendation performance. 

\begin{figure}
    \captionsetup[subfloat]
    {}
    \centering
    \subfloat[ZhihuRec]{
    \label{rq2-pop-zhihu}
    \includegraphics[width=0.48\linewidth,height=0.42\linewidth]{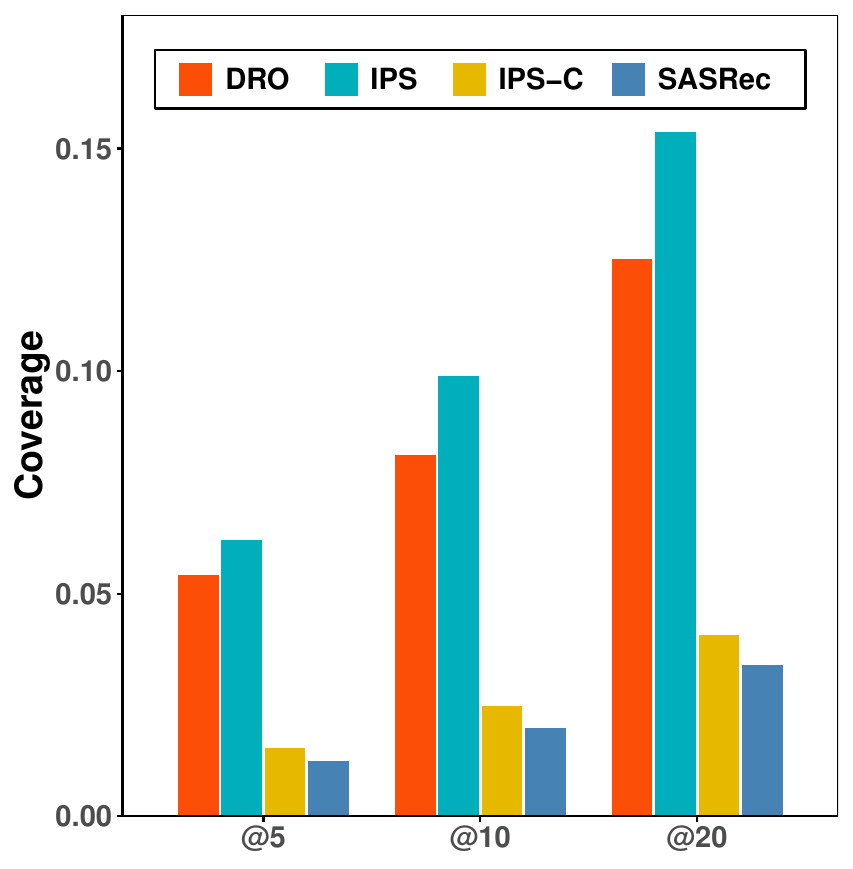}}
    \hspace{0.1cm}
    \subfloat[Tenrec]{%
    \label{rq2-pop-tenrec}
    \includegraphics[width=0.48\linewidth,height=0.42\linewidth]{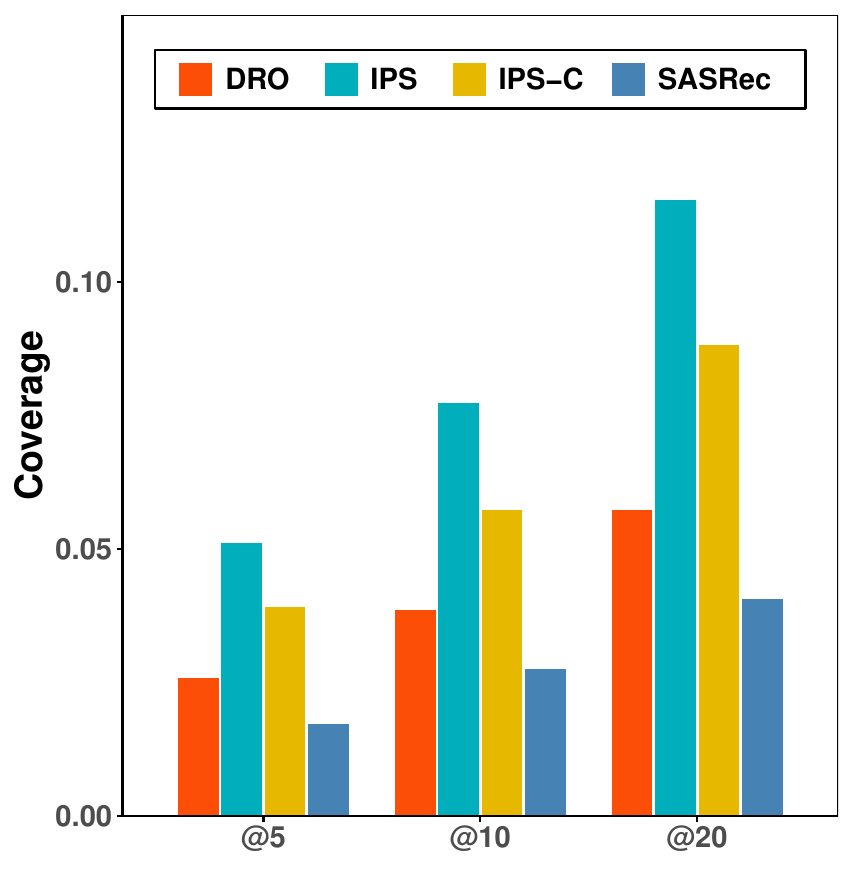}}
    \caption{Popularity debiasing effect with different methods.}
    \label{rq2-pop}
\end{figure}

Exposure bias can amplify popularity bias, resulting in relevant but unpopular items being overlooked and not recommended \cite{biassurvey, gupta2021correcting}. Therefore, we also conduct experiments to see whether the proposed framework generates more diverse recommendation.
We use SASRec as the backbone model.
The coverage metric:$\text { Coverage} @K =\frac{\left|\cup_{u \in \text { test }} list_{@K}(u)\right|}{|\mathcal{I}|}$,
where $list_{@K}(u)$ is the top-$K$ recommendation list for user $u$, is used to measure the recommendation diversity.
As shown in Fig. \ref{rq2-pop}, the reported results indicate that the proposed method can significantly help to alleviate popularity bias. 
We can see that the DRO, IPS, and IPS-C methods outperform the vanilla backbone model (SASRec) on both datasets. Specifically, the IPS method achieves the best performance.
However, as reported in Table \ref{table:rq1}, the accuracy of the IPS method cannot achieve satisfying results.
Compared with vanilla SASRec, the DRO method achieves better coverage performance, because the recommender trained by DRO tends to make a conservative recommendation regarding biased samples. More concretely, the DRO method tends to mitigate the risk of continuously recommending high-exposed items. It indicates that alleviating exposure bias could potentially reduce popularity bias as well. It should be noted that the exposure simulator includes the use of a popularity-based recommender to boost popular items, which may also contribute to the success of these methods in alleviating popularity bias.

To conclude, the proposed DRO can effectively alleviate both exposure and popularity bias, and meanwhile generate more accurate recommendation compared with existing IPS-based methods.

\subsection{Effect of the Exposure Simulator (RQ2)}
In this subsection, we conduct experiments to verify the impact of the exposure simulator on the Tenrec dataset. 
We first examine how the mixture of recommenders affects the modeling of exposure data. It is important to accurately simulate the exposure distribution since  DRO relies on the system exposure distribution to construct the uncertainty set.
Besides, the performance of IPS-based methods is also reliant on the propensity, which is the exposure probability in our setting. Table \ref{table:rq3-exposure} shows the simulator performance comparison. 
Notably, ``Expo-M'' denotes the exposure simulator that combines Expo-SASRec, Expo-GRU4Rec, and popularity-based recommenders as detailed in subsection \ref{subsec:model_exposure}. ``Expo-G''
is short for Expo-GRU4Rec and ``Expo-S'' is short for Expo-SASRec.
We can see that the ''Expo-M" simulator outperforms the other single model simulators, indicating its superior ability to accurately model the exposure distribution. 

Table \ref{table:rq3-debiasing} shows the debiasing performance comparison between IPS-C and the proposed DRO method with different exposure simulators. 
We can see that the debiasing performance with the ``Expo-M'' simulator achieves better performance than vanilla 
SASRec and that with the ``Expo-S'' simulator. It indicates the ``Expo-M'' simulator has remarkable ability to model exposure distribution and thus provides better debiasing performance. 
Moreover, the performance of the IPS-C method with the ``Expo-S'' simulator is inferior to that of the vanilla SASRec recommender. It shows that establishing a good exposure simulator for the propensity score is indispensable for IPS-based methods. 
\begin{table}
    \centering
    \setlength{\abovecaptionskip}{3pt}
    \begin{threeparttable}
    \caption{Performance of exposure simulator. Boldface denotes the highest score. ``R'' and ``N'' are short for Recall and NDCG. ``Expo-M'' denotes the exposure simulator based on mixture of recommenders. ``Expo-G'' and ``Expo-S'' are short for Expo-GRU4Rec and Expo-SASRec, respectively.} 
    \renewcommand\arraystretch{0.9}
    \label{table:rq3-exposure}
    \begin{tabular}{c|cc|cc|cc}
    \toprule
    Simulator&R@5&N@5&R@10&N@10&R@20&N@20\cr
    \midrule
    Expo-G& 0.0320& 0.0210& 0.0513& 0.0273& 0.0706& 0.0322\cr
    Expo-S& 0.0335& 0.0218& 0.0524& 0.0280& 0.0735& 0.0333\cr
    \midrule
    Expo-M& \textbf{0.0381}& \textbf{0.0244}& \textbf{0.0533}& \textbf{0.0293}& \textbf{0.0802}& \textbf{0.0361}\cr
    \bottomrule
    \end{tabular}
    \end{threeparttable}
\end{table}
\begin{table}
    \centering
    \setlength{\abovecaptionskip}{3pt}
    \begin{threeparttable}
    \caption{Debiasing performance $w.r.t$ different exposure simulators. Boldface denotes the highest score.  ``R'' and ``N'' are short for Recall and NDCG. ``Expo-M'' denotes the exposure simulator based on mixture of multiple recommenders. ``Expo-S'' is short for Expo-SASRec.}
    \renewcommand\arraystretch{0.9}
    \label{table:rq3-debiasing}
    \begin{tabular}{p{1cm}<{\centering}p{0.8cm}<{\centering}p{0.7cm}<{\centering}p{0.7cm}<{\centering}p{0.7cm}<{\centering}p{0.7cm}<{\centering}p{0.7cm}<{\centering}p{0.7cm}<{\centering}}
    \toprule
    Simulator&Method&R@5&N@5&R@10&N@10&R@20&N@20\cr
    \midrule
    -- &SASRec & 0.033& 0.021& 0.063& 0.031& 0.104& 0.041\cr
    \midrule
    \multirow{2}{*}{Expo-S}
    &IPS-C  & 0.027& 0.017& 0.047& 0.024& 0.079& 0.032\cr
    &DRO  & 0.044& 0.028& 0.082& 0.041& 0.129& 0.052\cr
    \midrule
    \multirow{2}{*}{Expo-M}
    &IPS-C  & 0.041& 0.026& 0.073& 0.037& 0.114& 0.047\cr
    &DRO  &\textbf{0.052}& \textbf{0.032}& \textbf{0.088}& \textbf{0.044}& \textbf{0.141}& \textbf{0.057}\cr
    \bottomrule
    \end{tabular}
    \end{threeparttable}
\end{table}
\subsection{Hyperparameter Study (RQ3)}
\begin{figure}[t]
    \captionsetup[subfloat]{justification=centering}
    \centering
    \subfloat[Recall@20, Tenrec]{
    \label{rq4-tenrec-recall}
    \includegraphics[width=0.45\linewidth,height=0.45\linewidth]{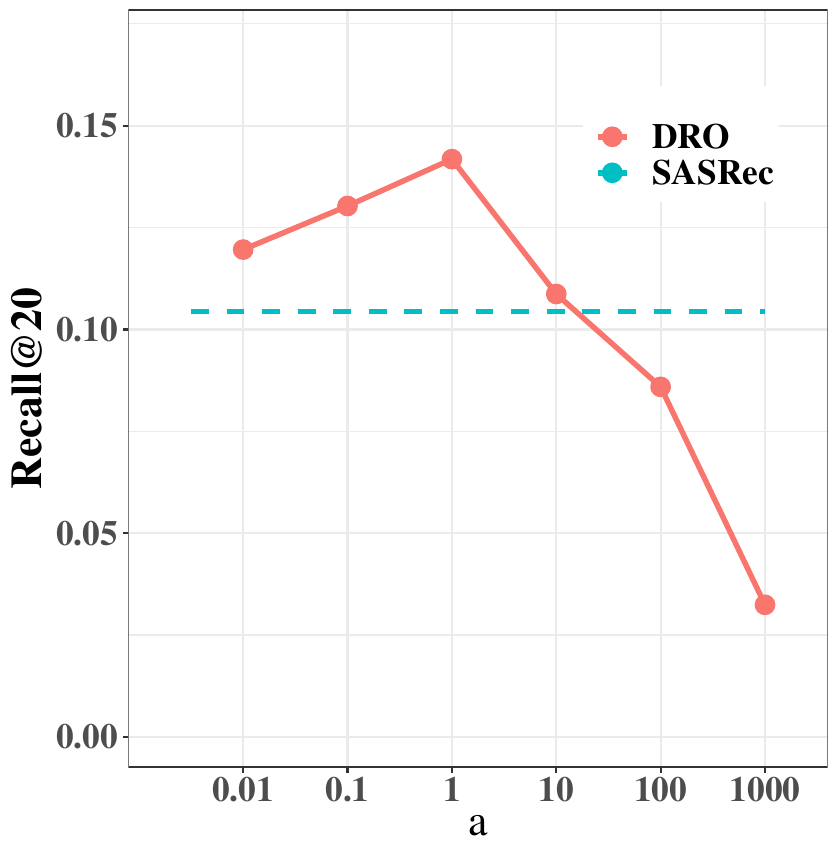}}
    \hspace{0.3cm}
    \subfloat[NDCG@20, Tenrec]{%
    \label{rq4-tenrec-ndcg}
    \includegraphics[width=0.45\linewidth,height=0.45\linewidth]{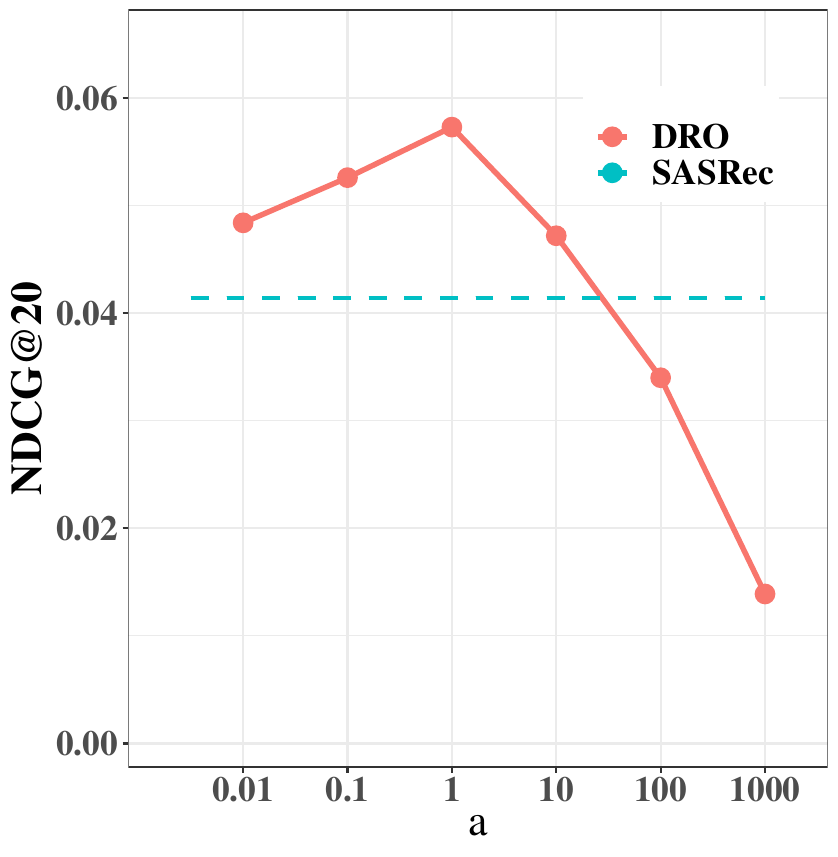}}
    \vfill
    \subfloat[Recall@20, ZhihuRec]{%
    \label{rq4-zhihurec-recall}
    \includegraphics[width=0.45\linewidth,height=0.45\linewidth]{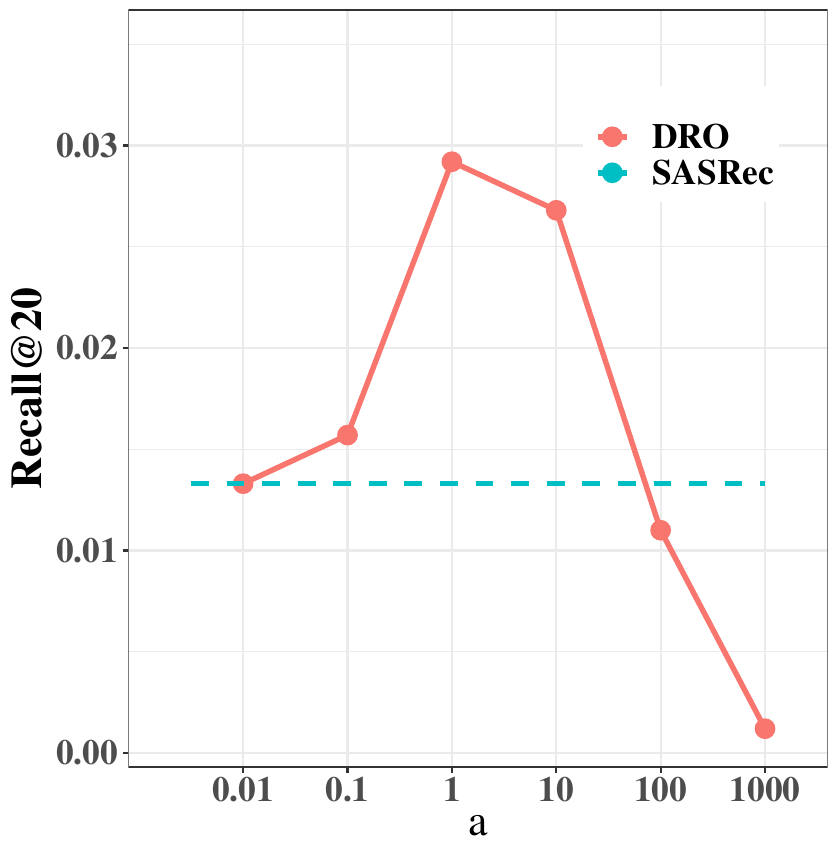}}
    \hspace{0.3cm}
    \subfloat[NDCG@20, ZhihuRec]{%
    \label{rq4-zhihurec-ndcg}
    \includegraphics[width=0.45\linewidth,height=0.45\linewidth]{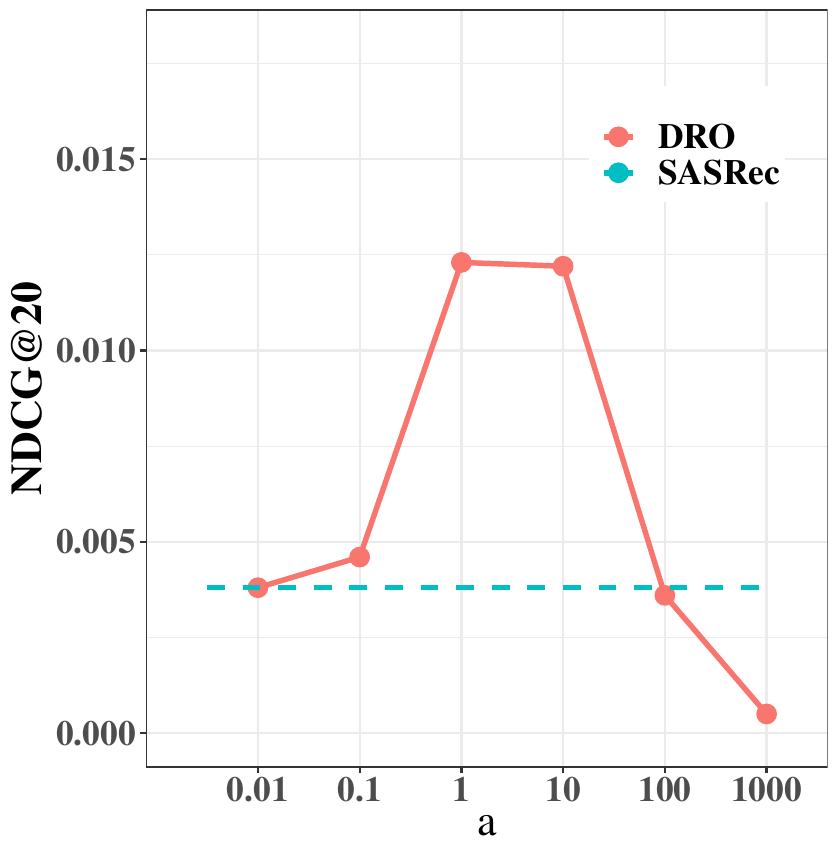}}
    \caption{
    Effect of the weight of the DRO loss. The dash line denotes the performance of SASRec.}
    \label{rq4-kstudy}
\end{figure}

In this subsection, we conduct experiments to demonstrate the impact of the different weights of the DRO loss. We use SASRec as the backbone recommendation model. Results on the GRU4Rec show the same trend. As discussed in subsection \ref{subsection:Over-estimation}, the recommender trained by joint optimization objective (as detailed in Eq.(\ref{eq:training_loss})) with the suitable value of $a$ may approximate the user's true preference meanwhile mitigate the exposure bias. If $a \rightarrow 0$, the optimization objective is reduced to the recommender loss, the recommender is solely trained on biased user interactions and is likely to overestimate the user preference over biased samples. If $a \rightarrow \infty$, the recommender would have excessive attention to perform debiasing and lose recommendation accuracy. Fig. \ref{rq4-kstudy} shows the performance of DRO with different $a$ tuned in the range of [0.01, 0.1, 1, 10, 100, 1000] on both datasets.

As shown in Fig. \ref{rq4-tenrec-recall} and \ref{rq4-tenrec-ndcg}, we can observe the performance of DRO with $a$ in [0.01, 0.1, 1,10] outperforms the SASRec on the Tenrec dataset. The results show that DRO with suitable $a$ can perform exposure debiasing and better obtain user preference compared with the backbone model SASRec. Regarding the DRO with $a$ in [100, 1000], the recommendation performance drops dramatically. The reason is that the recommender neglects user personalization when it over-focuses on debiasing. The results on ZhihuRec are basically the same as those in Tenrec. Different from the Tenrec, the performance of DRO with $a=0.01$ is comparable with the SASRec on the ZhihuRec dataset.
\section{Conclusions and Future Work}
\label{sec:conclusion}
In this paper, we have investigated the task of exposure debiasing in sequential recommendation.
We have proposed to adopt Distributionally Robust Optimization (DRO) to perform exposure debiasing in SR. More specifically, we have introduced an exposure simulator trained upon the system exposure data to calculate exposure distribution, which is then regarded as the nominal distribution to construct the uncertainty set of DRO. Besides, we have introduced the penalty to items with high exposure probability to avoid the overestimation issue for the preference scores of biased samples. 
Furthermore, we have designed a debiased self-normalized inverse propensity score (SNIPS) evaluator for evaluating the debiasing sequential recommendation performance in the biased test set.
Extensive experiments and analysis on two real-world datasets have demonstrated the effectiveness of the proposed framework.

In future work, we plan to design more advanced exposure simulators to better model the system exposure distribution. In the real-world recommendation scenario, items are exposed page-by-page. To this end, developing exposure simulators to model page-wise exposure data would be a potential research direction. Additionally, the system exposure data is also highly affected by other kinds of biases (e.g., the popularity bias).  Exploring the relationship between exposure bias and other biases would also be an interesting research direction. We hope this work
can raise more community concern to conduct recommendation debiasing from a more wide range of perspectives, other than just focusing on the user interaction data.

\section*{Acknowledgements}
This work was supported by the National Key R\&D Program of China with grants No. 2020YFB1406704 and No. 2022YFC3303004, the Natural Science Foundation of China (62202271, 61902219, 61972234, 61672324, 62072279, 62102234, 62272274), the Natural Science Foundation of Shandong Province (ZR2021QF129, ZR2022QF004), the Key Scientific and Technological Innovation Program of Shandong Province (2019JZZY010129), the Fundamental Research Funds of Shandong University, the Tencent WeChat Rhino-Bird Focused Research Program (WXG-FR-2023-07). All content represents the opinion of the authors, which is not necessarily shared or endorsed by their respective employers and/or sponsors. 
\clearpage
\bibliographystyle{ACM-Reference-Format}
\balance
\bibliography{main}
\clearpage
\appendix
\begin{appendices}
\section{PROOF OF THEOREMS}
\label{sec:appendix_proof}
\subsection{Proof 1}
\label{subsec:appendix_proof_dro}
\begin{proof} In this proof, we follow the deviation from \cite{hu2013kullback}. Set $D$ as KL-divergence, minimizing the Eq. (\ref{eq:dro1}) can be formulated as:
\begin{equation}
\begin{split}
    \min_{\theta} \max_{q} \mathbb{E}_{(S^u, v) \sim q} \ell(S^u, v, \theta) \\
    \text{subject to } D_{KL}\left(q \| q_{0}\right) \leq \eta.
    \label{dro_appendix_init}
\end{split}
\end{equation}
Note that $q_0$ is the nominal distribution. Assume that the distributions $q$ and $q_0$ have densities denoted by $p(S^u, v)$ and $p_0(S^u, v)$, respectively. 

Let the Radon-Nikodym derivative $L(S^u, v) = p(S^u, v)/p_0(S^u, v)$, and by applying the change-of-measure technique, we obtain:
\begin{equation}
    \begin{split}
    D_{KL}\left(q \| q_{0}\right) 
    &= \int \frac{p(S^u, v)}{p_0(S^u, v)} \log \frac{p(S^u, v)}{p_0(S^u, v)} p_0(S^u, v) \mathrm{d} s \mathrm{d} v \\
    &= \int L(S^u, v) \log L(S^u, v) p_0(S^u, v) \mathrm{d} s \mathrm{d} v \\
    &= \mathbb{E}_{(S^u, v) \sim q_0}[L(S^u, v) \log L(S^u, v)].
    \end{split}
\end{equation} 
Similarly, 
\begin{equation}
    \mathbb{E}_{(S^u, v) \sim q} \ell(S^u, v, \theta) = 
    \mathbb{E}_{(S^u, v) \sim q_0}[\ell(S^u, v, \theta) L(S^u, v)].
\end{equation} 
Then, the inner maximization problem in Eq.(\ref{dro_appendix_init}) can be reformulated as
\begin{equation}
\begin{split}
    \max_{L} \mathbb{E}_{(S^u, v) \sim q_0}[\ell(S^u, v, \theta) L(S^u, v)] \\
    \text{subject to } \mathbb{E}_{(S^u, v) \sim q_0} [L\log L] \leq \eta. 
    \end{split}
    \label{dro_appendix_change_measure}
\end{equation}
To formulate the dual problem of Eq.(\ref{dro_appendix_change_measure}), we formulate the Lagrangian functional is 
\begin{equation}
    l(\alpha,L) = \mathbb{E}_{(S^u, v) \sim q_0}[\ell(S^u, v, \theta) L] - \alpha(\mathbb{E}_{(S^u, v) \sim q_0} [L\log L] - \eta)
    \label{lagra_fun}
\end{equation}
Then, Eq.(\ref{dro_appendix_change_measure}) is equivalent to
\begin{equation}
    \max_{L} \min_{\alpha \geq 0}  l(\alpha,L).
    \label{dro_lagra}
\end{equation}
Interchanging the order of the maximum and minimum operators, we obtain the Lagrangian dual of Eq.(\ref{dro_lagra}), which is represented as
\begin{equation}
    \min_{\alpha \geq 0} \max_{L} l(\alpha,L).
    \label{dro_lagra_dual}
\end{equation}
The strong duality for Eq.(\ref{dro_lagra}) and Eq.(\ref{dro_lagra_dual}) have proved in \cite{hu2013kullback}.
It is easy to see the $\mathbb{E}_{(S^u, v) \sim q_0}[L]=1$.
Omitting the term $\alpha\eta$, the inner maximization problem in Eq.(\ref{dro_lagra_dual}) can be expressed as
\begin{equation}
\begin{split}
    \max_{L} \mathbb{E}_{(S^u, v) \sim q_0}[\ell(S^u, v, \theta) L - \alpha L\log L]) \\
    \text{subject to } \mathbb{E}_{(S^u, v) \sim q_0}[L] = 1.
    \label{close_form}
\end{split}
\end{equation}
Define the functionals of Eq.(\ref{close_form})
\begin{equation}
    \begin{split}
        \mathcal{J}(L) &= \mathbb{E}_{(S^u, v) \sim q_0}[\ell(S^u, v, \theta) L - \alpha L\log L]) \\
        \mathcal{J}_{c}(L) &= \mathbb{E}_{(S^u, v) \sim q_0}[L] - 1.
    \end{split}
    \label{functional}
\end{equation}

Let $D \mathcal{J}(L(S^u, v))$ denote the derivative of $\mathcal{J}(L(S^u, v))$, and for any feasible direction $V(S^u, v)$ at $L(S^u, v)$, we have:
\begin{flalign}
    \label{functional_derivative}
    &D \mathcal{J}(L(S^u, v)) V(S^u, v) \nonumber\\
    =&\lim _{t \rightarrow 0} \frac{\mathcal{J}(L(S^u, v)+t V(S^u, v))-\mathcal{J}(L(S^u, v))}{t} \nonumber\\
    =&\lim _{t \rightarrow 0} \frac{\mathbb{E}_{(S^u, v) \sim q_0}[\ell(S^u, v, \theta)(L+t V)-\alpha(L+t V) \log (L+t V)]}{t} \nonumber\\
    -& \lim _{t \rightarrow 0} \frac{\mathbb{E}_{(S^u, v) \sim q_0}[\ell(S^u, v, \theta) L-\alpha L \log L]}{t} \nonumber \\
    &=\mathbb{E}_{(S^u, v) \sim q_0}[\ell(S^u, v, \theta) V] \nonumber \\
    -&\alpha \lim _{t \rightarrow 0} \mathbb{E}_{(S^u, v) \sim q_0}\left[\frac{(L+t V) \log (L+t V)-L \log L}{t}\right].
\end{flalign}
Using the monotone convergence theorem, we can rearrange the operators in Eq.(\ref{functional_derivative}): 
\begin{flalign}
&D \mathcal{J}(L(S^u, v)) V(S^u, v) \nonumber\\ 
& =\mathbb{E}_{(S^u, v) \sim q_0}[\ell(S^u, v, \theta) V] \nonumber \\
& -\alpha \mathbb{E}_{(S^u, v) \sim q_0} \lim_{t \rightarrow 0}\left[\frac{(L+t V) \log (L+t V)-L \log L}{t}\right]\nonumber \\
& =\mathbb{E}_{(S^u, v) \sim q_0}[(\ell(S^u, v, \theta) - \alpha(\log L + 1))V].
\end{flalign}
Similarly, we have: 
\begin{equation}
   D \mathcal{J}_c(L(S^u, v)) V = \mathbb{E}_{(S^u, v) \sim q_0}[V].
\end{equation}
We construct Lagrangian functional associated with Eq.(\ref{lagra_fun}) as follows:
\begin{equation}
\begin{aligned}
  Y(L,\lambda) &= \mathbb{E}_{(S^u, v) \sim q_0}[\ell(S^u, v, \theta) L - \alpha L\log L] + \lambda(\mathbb{E}_{(S^u, v) \sim q_0}[L] - 1) \\
  & = \mathbb{E}_{(S^u, v) \sim q_0}[\ell(S^u, v, \theta) L - \alpha L\log L + \lambda L] - \lambda.
\end{aligned}
\end{equation}
Similarly, it is also straightforward to obtain that
\begin{equation}
\begin{aligned}
D Y(S^u, v) V
=\mathbb{E}_{(S^u, v) \sim q_0}[(\ell(S^u, v, \theta) - \alpha(\log L + 1)+ \lambda)V].
\end{aligned}
\end{equation}
To acquire an optimal solution $L^*(S^u, v)$, let $D Y(L(S^u, v))=0$ for the all feasible direction $V(S^u, v)$:
\begin{equation}
    \ell(S^u, v, \theta) - \alpha(\log L + 1)+ \lambda = 0.
\end{equation}
Solving the equation, we get 
\begin{equation}
    L^*(s,v,\lambda)= e^{\ell(S^u, v, \theta)/\alpha +(\lambda-\alpha)/\alpha}.
\end{equation}
From Eq.(\ref{close_form}) and Eq.(\ref{functional}) we have $\mathbb{E}_{(S^u, v) \sim q_0}[L^*] = 1$ and $\mathcal{J}_{c}(L^*)=0$, then we can obtain $\lambda^* = -\alpha \mathbb{E}_{(S^u, v) \sim q_0}[e^{\ell(S^u, v, \theta)/\alpha}] + \alpha$.
Therefore, 
\begin{equation}
    L^*(S^u, v) = L^*(s,v,\lambda^*) = \frac{e^{\ell(S^u, v, \theta)/\alpha}}{\mathbb{E}_{(S^u, v) \sim q_0}[e^{\ell(S^u, v, \theta)/\alpha}]}
\end{equation}
Put $L^*(S^u, v)$ into Eq.(\ref{lagra_fun}), we obtain the close from of Eq.(\ref{dro_appendix_change_measure}):
\begin{equation}
    l^*(\alpha,L)=\alpha \mathbb{E}_{(S^u, v) \sim q_0}[e^{\ell(S^u, v, \theta)/\alpha}] + \alpha \eta
\end{equation}
Finally, the $\alpha =1 $ in our setting and omitting constant terms, we can obtain:
\begin{equation}
    \mathcal{L}_{DRO}^{''}(\theta) = \mathbb{E}_{(S^u, v) \sim q_0}[e^{\ell(S^u, v, \theta)}],
\end{equation}
which completes the proof.
\end{proof}
\subsection{Proof 2}
\label{subsec:appendix_proof_snips}
\begin{proof}
In this proof, we follow \cite{Yang2018unbiasevaluatuion} to prove the following SNIPS evaluator is a debiased evaluator.
\begin{equation}
    \begin{split}
    \label{eq:E_SNIPS)_proof}
    \mathbb{E}_O[\hat{R}_{\text{SNIPS}}(\hat{Z}|\rho)] &= \frac{1}{\sum_{u \in \mathcal{U}}{\frac{1}{\rho^{k}_{v}}}} \sum_{u \in \mathcal{U}} \sum_{i \in \mathcal{I}} \frac{c(\hat{Z}_{u, \, i)}}{\rho^{k}_{i}} \cdot \mathbb{E}_O[O_{u,i}] \\
    &= \frac{1}{\sum_{u \in \mathcal{U}}{\frac{1}{\rho^{k}_{v}}}} \sum_{u \in \mathcal{U}} \sum_{i \in \mathcal{I}} \frac{c(\hat{Z}_{u, \, i)} }{\rho^{k}_{i}} \cdot \rho_{i} \\
    &= \frac{1}{\sum_{u \in \mathcal{U}}{\frac{1}{\rho^{k}_{v}}}} \sum_{u \in \mathcal{U}} \sum_{i \in \mathcal{I}} \frac{c(\hat{Z}_{u, \, i)}}{\rho^{1-k}_{i}} \\
    &= \frac{|\mathcal{U}| \cdot |\mathcal{I}|}{\sum_{u \in \mathcal{U}}{\frac{1}{\rho^{k}_{v}}}} \sum_{u \in \mathcal{U}} \sum_{i \in \mathcal{I}} \frac{1}{\rho^{1-k}_{i}} \cdot \mathbb{E}_O[R_{\text{ideal}}(\hat{Z})]. 
    \end{split}
\end{equation}
\end{proof}
\vspace{-0.4cm}
\end{appendices}

\begin{appendices}
\section{statistics of datasets}
\label{sec:appendix_dataset}
In this work, we conduct experiments on two datasets, Zhihu~\citep{hao2021largescale} and Tenrec ~\citep{tenrec}. Table \ref{Datasets} summarizes the statistics of the two datasets. 

\header{ZhihuRec}. This dataset is obtained from Zhihu, a knowledge-sharing platform. The original data includes question information, answer information, and user profiles. Our focus is on the scenario of answer recommendation, where a slate of answers (i.e., the items in our setting) is presented to the user during the recommender's serving period.
The dataset contains the show time and click time of all answers, with a value of 0 for non-clicked answers.
In total, the dataset contains 771,550 impressions and 214,853 user clicks, covering 9,223 items from 7,862 users. 

\header{Tenrec}. This is a large-scale dataset and is collected from two feeds, namely articles and videos, on Tenrec's recommendation platforms. Our primary focus for this study is on the video recommendation scenario. The data we used comprises 1,738,116 recommended impressions for 31,722 users, covering 24,653 videos and a total of 912,812 user clicks.
\begin{table}[!t]
    \centering
    \setlength{\abovecaptionskip}{3pt}
    \begin{threeparttable}
    \caption{Dataset statistics.}
    \label{Datasets}
    \begin{tabular}{cccc}
    \toprule
    Dataset  & ZhihuRec & Tenrec\cr
    \midrule
    \#users  &7,862 & 31,722\cr
    \#items  &9,223 & 24,653\cr
    \#impressions  &771,550 &1,738,116\cr
    \#clicks  &214,853 &912,812\cr
    \bottomrule
  \end{tabular}
    \end{threeparttable}
\end{table}
\end{appendices}

\begin{appendices}

\section{reproducibility}
\label{appedix:reproducibility}
\header{Hyperparameter settings.}
For each user interaction, we preserve the last 50 click items as the historical interaction sequence to avoid too long sequences. We pad the sequence with an additional padding item if the sequence length is less than 50. For the system exposure, we preserve the last 200 exposed items and pad the sequence with an additional padding item if the sequence length is less than 200.
For a fair comparison, the item embedding size is set to 64 across all models. 
We train all models with the Adam optimizer\cite{DBLP:journals/corr/KingmaB14}. The learning rate is tuned to 0.005.
For SASRec, the number of heads in self-attention is set to 2, with a total of 64 hidden neurons. 
For GRU4Rec, the number of layers in GRU is set to 1. The size of hidden layers in the feedforward network is also set to 64 for two backbone models. 
The $\beta$ in Eq.(\ref{eq:predict}) to boost popular items is set to 0.3. Each experiment is conducted 3 times and the average result is reported.
\end{appendices}

\end{document}